# $V_{Br}$ >10 kV E-Beam/Sputtered Vertical $NiO_x$/(011) β-$Ga_2O_3$ HJDs with PFOM >2.3 $GW/cm^2$

Y. Liu[1], C. Peterson[1], C. N. Saha[1], M. J. Tadjer[2], and S. Krishnamoorthy[1]
[1]Materials Department, University of California, Santa Barbara, Santa Barbara, CA, U.S.A.
[2]United States Naval Research Laboratory (NRL), Washington D.C., U.S.A.
*Electronic mail:* yizhengliu@ucsb.edu, sriramkrishnamoorthy@ucsb.edu

***Abstract*—**Beta-gallium oxide (β-$Ga_2O_3$) holds enormous potential for medium voltage range power electronic applications. This work reports $V_{Br}$ > 10 kV/$R_{on,sp}$ = 43 mΩ•$cm^2$ class edge terminated vertical heterojunction diodes (HJDs) with e-beam/sputtered nickel oxide ($NiO_x$) stack on epitaxial (011) β-$Ga_2O_3$. The power figure of merit (PFOM) of the HJD exceeds 2.3 $GW/cm^2$. The extracted parallel plane breakdown field is > 5.3 MV/cm, which is the highest reported electric field for thick (011) β-$Ga_2O_3$ epitaxial drift layer.

## I. INTRODUCTION

The rapid expansion of data centers for artificial intelligence (AI) and charging networks for electric vehicles (EVs) have stimulated the necessity for efficient, cost-effective grid-level transmission and power conversion circuits at medium voltage level (1-35 kV).[1] Solid states transformers (SSTs) are critical for data centers in enabling efficient, high power density distribution by replacing conventional magnetic transformers with power semiconductor-based converters. Enormous market potential is available in direct-to-rack power conversion for AI data centers that facilitate large voltage (13.5-35 kV, AC) to 800 V DC to reduce immediate power conversion steps, improving efficiencies for AI graphics processing units (GPUs) and accelerators. The high critical electric field strength (6-8 MV/cm) and availability of the shallow hydrogenic dopants[2] in epitaxial β-$Ga_2O_3$[3], [4], [5], [6] can be leveraged to demonstrate power devices with much lower differential specific on-resistance while operating at a higher blocking voltage[7], [8], [9] compared to silicon carbide (SiC) and gallium nitride (GaN)[10], [11]. Although reliable p-type doping is currently not available in β-$Ga_2O_3$, heterojunctions formed with p-$NiO_x$ are able to realize multi-kV class[12] junctions along with breakdown electric field anisotropy across various orientations.[13], [14]

In this work, we demonstrated a >10 kV $NiO_x$-based (e-beam/sputtered stack) vertical HJD with 43 mΩ•$cm^2$ $R_{on,sp}$ and PFOM value > 2.3 $GW/cm^2$ on (011) homoepitaxial β-$Ga_2O_3$ with high electric field handling capability. [14] The parallel-plane breakdown electric field is extracted to be > 5.3 MV/cm.

## II. DEVICE DESIGN AND FABRICATION

The $NiO_x$/(011) HVPE β-$Ga_2O_3$ HJD fabrication begins with a backside Ti/Au (50/350 nm) Ohmic metallization on $n^+$ β-$Ga_2O_3$ bulk substrate using e-beam evaporation followed by a 60-seconds rapid thermal annealing (RTA) at 470 °C in $N_2$. An 8 nm of e-beam-evaporated $NiO_x$ is deposited on (011) HVPE β-$Ga_2O_3$ drift region (Novel Crystal Technology (NCT), ~20 μm, $2\times10^{15}$ $cm^{-3}$) with a pre-patterned photoresist liftoff mask by optical lithography to circumvent the ion channeling effect in the (011) crystal orientation[15]. Following the first layer $NiO_x$ deposition, a self-aligned $p^-$ $NiO_x$ (~20 nm) and $p^{++}$ $NiO_x$ (~20 nm) contact layer are reactively sputtered. Then, a Ni/Au/Ni (50/100/150 nm) anode cap/metal hard mask stack is deposited via e-beam evaporation. The fabricated HJDs are later dry-etched ~2 μm deep into the β-$Ga_2O_3$ drift region below the heterojunction interface using a $BCl_3$ inductively coupled plasma (ICP) process at 200 W to accomplish mesa isolation with rounded corners for edge termination[16], [17], mitigating electric field crowding at device edges. A ~1.5 μm-thick $SiO_2$ field plate oxide is sputtered conformally on the mesa-isolated (MI) HJDs, then a Ni/Au (50/300 nm) field plate (FP) metal stack a with 20-μm edge extension ($L_{FP}$) is deposited on the oxide via e-beam evaporation to conclude the device fabrication. Cross-sectional device schematics is shown in **<u>Fig. 1.(a)</u>** for Schottky barrier diode (SBD) reference and **<u>Fig. 1.(b)</u>** for MI/FP HJDs.

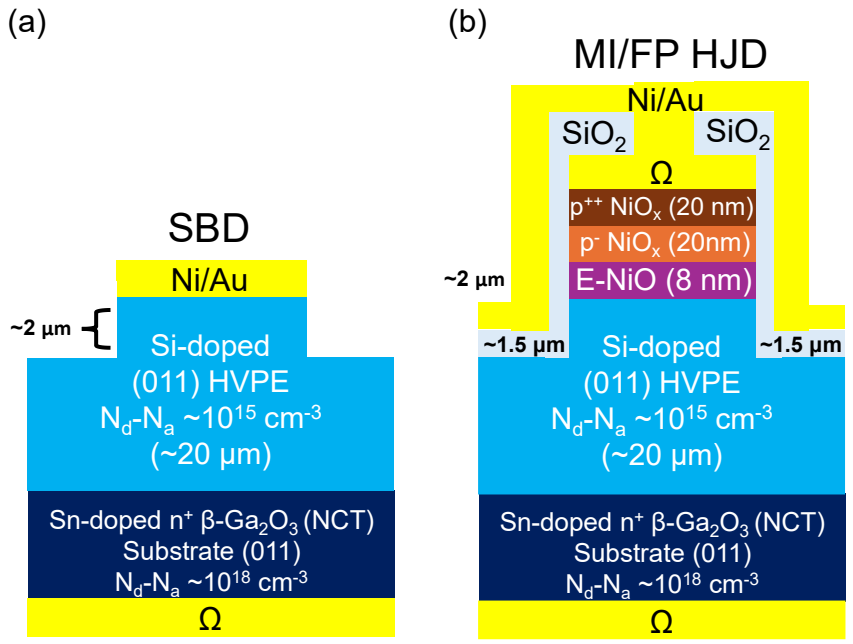

Fig.1. Schematics cross-section of (a) reference Schottky barrier diodes (SBDs) and (b) mesa-isolated/field-plated heterojunction diodes (MI/FP HJDs).

## III. DEVICE CHARACTERIZATION

The electrical characteristics of the device are measured on Keithley 4200 Parameter Analyzer. The forward linear current-voltage (J-V) characteristics of the MI/FP HJDs are shown in **Fig.2. (a)** on circular diodes with device dimensions from 60-µm dia. to 1-mm dia. along with those of SBD counterparts. The MI/FP HJDs exhibit on-state current density at 60-80 A/cm$^2$ at 5 V with a turn-on voltage at 2.2 V across different device sizes. The SBDs show forward current density 80-90 A/cm$^2$ at the same forward bias. Semi-log scale J-V characteristics of MI/FP HJDs in **Fig.2. (b)** reveal the reverse leakage current density at $10^{-9}$-$10^{-8}$ A/cm$^2$ under -5 V reverse voltage with a rectifying ratio at $10^{10}$-$10^{11}$ across various diode dimensions. **Fig.2. (c)** shows that the HJDs' differential specific on-resistances ($R_{on,sp}$) with various device areas match consistently with each other at 38-43 mΩ•cm$^2$ if considering a non-current spreading model, whereas they differ significantly if a 45°-angle current spreading[17] model is considered. Therefore, the non-current spreading model is adopted for all devices J-V characteristics normalization. The pulsed I-V characteristics with <1% duty cycle in **Fig.2. (d)** of the MI/FP HJD on 1-mm dia. device accomplishes > 1.5 A absolute current at 9 V on low-doped and thick HVPE grown (011) β-$Ga_2O_3$. The current value of the HJD closely matches that of the SBD counterpart with same device dimension, indicating that the e-beam/sputtered MI/FP HJD fully exploits the on-state potential of the epitaxial drift region.

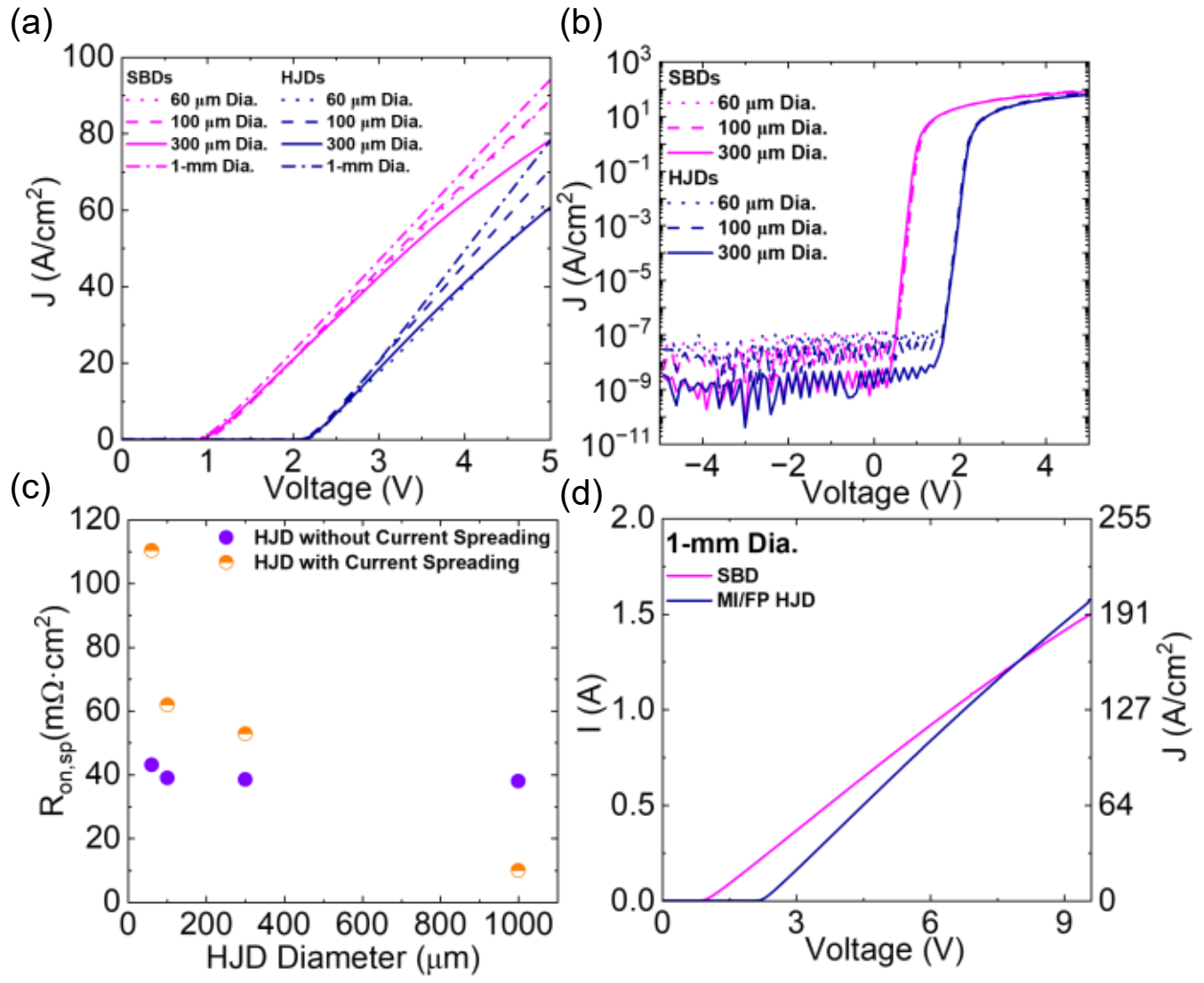

Fig.2. (a) Linear forward J-V characteristics of SBDs and MI/FP HJDs. (b) Semi-log scale J-V characteristics of SBDs and MI/FP HJDs. (c) $R_{on,sp}$ vs. diode dimension for SBDs and MI/FP HJDs. (d) Pulsed I-V characteristics for SBD and MI/FP 1-mm dia. devices.

Beyond J-V characterization, the capacitance-voltage (C-V) characteristics of the MI/FP HJD are shown in **Fig.3. (a)** on 1-mm dia. area at 1MHz. The inset of **Fig.3. (a)** linearly extracts the built-in ($V_{bi}$) potential of the HJD at 2.2 V via 1/C$^2$ vs. voltage characteristics consistent with the diode's turn-on voltage.

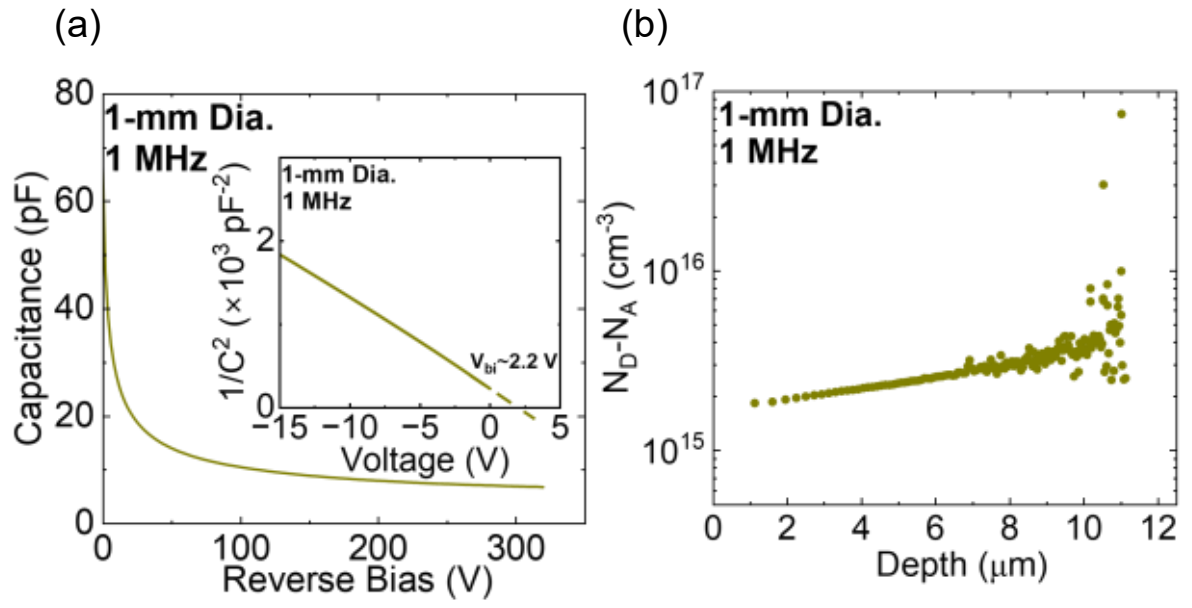

Fig.3. (a) C-V characteristics of MI/FP HJD on 1-mm dia. dimeison at 1 MHz with 1/C$^2$ vs. voltage inset. (b) Doping vs. depth profile MI/FP HJD on 1-mm dia. dimeison at 1 MHz.

The average apparent charge density of the (011) HVPE β-$Ga_2O_3$ is extracted to be $1.8\times10^{15}$ cm$^{-3}$ up to 11-12 µm, as shown in **Fig.3. (b)**, by using the relative permittivity value ($\varepsilon_s$) 10.79[14] for (011) β-$Ga_2O_3$.

The breakdown characteristics of the MI/FP HJDs are measured with a B1505A Semiconductor Parameter Analyzer. The 60-µm dia. HJDs showcases the highest breakdown voltage >10 kV and a lower bound catastrophic breakdown voltage at ~6 kV, as shown in **Fig. 4.(a).**

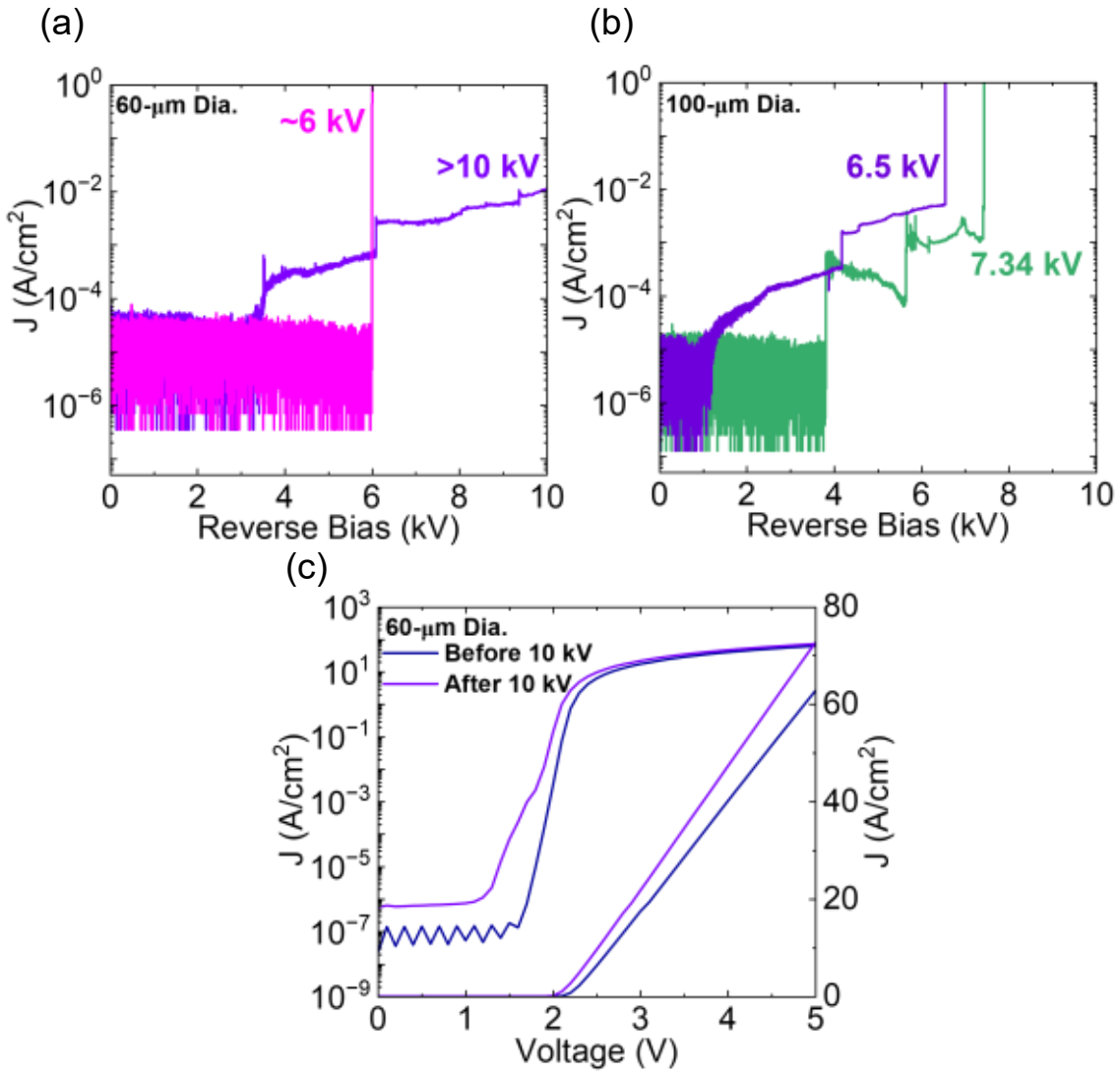

Fig.4. (a) Reverse breakdown characteristics of MI/FP HJDs for 60-µm dia. devices. (b) Reverse breakdown characteristics of MI/FP HJDs for 100-µm dia. devices. (c) J-V characteristics of 60-µm dia. device before and after 10 kV reverse bias in both linear and semi-log scale.

Additional MI/FP HJDs with 100-µm dia. device dimensions breaks down at 6.5-7.34 kV, as shown in **Fig. 4.(b).** The 60-µm dia. device which survived from the 10 kV reverse bias measurement is taken again to compare its J-V characteristics to that obtained before 10 kV sweep. The surviving device still exhibits rectifying behavior in semi-log scale with similar on-state current density at 60-70

A/cm$^2$ at 5 V forward voltage compared to its original J-V characteristics before experiencing 10 kV reverse bias, as shown in **Fig. 4.(c).** The slight rise of the reverse leakage current density and non-ideal turn-on behavior after 10 kV sweep can be possibly attributed to leakage and charge trapping in the field plate oxide.

**Fig. 5.** benchmarks the $R_{on,sp}$ vs. breakdown voltage ($V_{Br}$) of heterojunction diodes fabricated in this work in comparison to various HJDs reported in the literatures. It is shown that the PFOM of the MI/FP HJDs begins exceeding the material limit of 4H-SiC at 10 kV reverse bias.

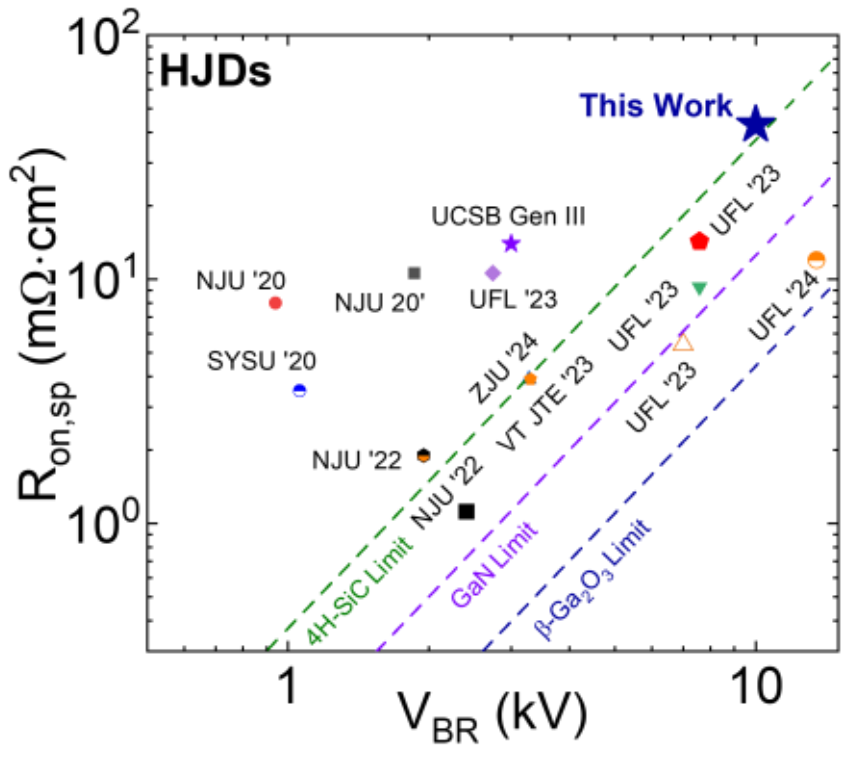

Fig.5. Benchmarking of $R_{on,sp}$ vs. breakdown voltage for various heterojunction diodes reported in the literature. [12], [16], [17], [19], [20], [21], [22], [23], [24], [25]

## IV. SUMMARY

This work demonstrates >10 kV class $NiO_x$/(011) HVPE β-$Ga_2O_3$ heterojunction diodes by combining e-beam and sputtered $NiO_x$ as the p-type layers. The field-plated HJD with mesa-isolation accomplishes a PFOM value > 2.3 GW/cm$^2$ and a parallel-plane breakdown electric field > 5.3 MV/cm. The high breakdown voltage/electric field accomplished on (011) β-$Ga_2O_3$ along with large PFOM value suggests this substrate orientation possesses potential for high electric field handling capability.

## ACKNOWLEDGEMENT

The author would like to acknowledge the funding from U.S. Department of Energy (DOE) ARPA-E ULTRAFAST program (DE-AR0001824) and Coherent/II-VI Foundation Block Gift Program. Research at NRL was partially supported by the Office of Naval Research.

## RFERENCES